\documentclass[a4paper,10pt]{article}
\usepackage{epsfig}
\usepackage{amssymb}
\begin{document}
\thispagestyle{empty}

\newcommand{\etal}  {{\it{et al.}}}  
\def\Journal#1#2#3#4{{#1} {\bf #2}, #3 (#4)}
\def\PRD{Phys.\ Rev.\ D}
\def\NIMA{Nucl.\ Instrum.\ Methods A}
\def\PRL{Phys.\ Rev.\ Lett.\ }
\def\PLB{Phys.\ Lett.\ B}
\def\EPJ{Eur.\ Phys.\ J}
\def\IEEETNS{IEEE Trans.\ Nucl.\ Sci.\ }
\def\CPCD{Comput.\ Phys.\ Commun.\ }

\smallskip

\bigskip
\bigskip

{\huge\bf
\begin{center}
Bose-Einstein correlations of neutral gauge bosons in $pp$ collisions
\end{center}
}

\vspace*{\fill}

\begin{center}
{\LARGE
 G.A. Kozlov

 }
\end{center}

\vspace*{\fill}

\begin{center}
\noindent
 {\large
 Bogolyubov Laboratory of Theoretical Physics\\
 Joint Institute for Nuclear Research,\\
 Joliot Curie st., 6, Dubna, Moscow region, 141980 Russia\\

 }
\end{center}
\vspace*{\fill}

 \section*{Abstract}
\noindent The theory for  Bose-Einstein correlations in case of
neutral gauge bosons in $pp$ collisions at high energies is presented.
Based on quantum field theory at finite temperature  the two-particle Bose-Einstein
correlations of neutral gauge bosons are carried out for the first time.
As a result, the important parameters of the correlation functions can be obtained for
the $Z^{0}Z^{0}$ pairs. The correlations of two bosons in 4-momentum space presented
in this paper offer useful
and instructive complimentary viewpoints to theoretical and experimental works in
multiparticle femtoscopy and interferometry measurements at hadron colliders.

\vspace*{\fill}


\newpage
\section{Introduction}

\bigskip

An investigation of the space-time extension or even squeezing of particle sources via the multiparticle
quantum-statistics correlation in high energy interactions is still attract the
attention of physical society in both experiment and theory. Over the past few decades, a considerable number of
successful studies have been done in this direction [1].
It is well
understood that the studies of correlations between produced
particles, the effects of coherence and chaoticity, an estimation of
particle emitting source size play an important role in high energy
physics.

By studying the Bose-Einstein correlations (BEC) of identical
particles (we mean like-sign charge particles and the neutral charge ones), it is possible  experimentally
to determine the time scale and spatial region over which particles do not
have the interactions. Such a surface is called as decoupling one.
In fact, for an evolving system such as $p p$ collisions, it is
not really a surface, since at each time there is a spread out
surface due to fluctuations in the last interactions, and the shape
of this surface evolve even in time. The particle source is not
approximately constant because of energy-momentum conservation
constraint.

More than half a century ago Hanbury-Brown and Twiss [2] used
BEC between photons to measure the size of distant stars. In the
works [3,4], the master equations for evolution of
thermodynamic system that can be created at the final state of
a high multiplicity process were established. The equations
have the form of the field operator evolution equation
(Langevin-like [5]) and allows one to gain  the basic
features of the emitting source space-time structure. In particular,
it has been conjectured and further confirmed that the size of the
emitting source through BEC is strongly affected by non-classical
off-shell effect.

The shapes of BEC function were experimentally established in the
LEP experiments ALEPH [6], DELPHI [7] and OPAL [8], and ZEUS Collaboration
at HERA [9], which
also indicated a dependence of the measured correlation radius on
the hadron $(\pi,\ K)$ mass. The results for $\pi^{\pm}\pi^{\pm}$ and
$\pi^{\pm}\pi^{\mp}$ correlations with $p\bar p$ collisions at $\sqrt {s}$ = 1.8 TeV
were published by E735 Collaboration in [10].

The correlations between heavy particles (e.g., neutral gauge
bosons $Z^{0}Z^{0}$) of Bose-Einstein type have not been carried
out previously at hadron colliders. Such a study can be addressed
to the Large Hadron Collider (LHC) which provide proton-proton interactions at
$\sqrt{s} = 14$ TeV centre-of-mass system (c.m.s.) energy.

In this work, we make an attempt to demonstrate that the problem
of properties of the genuine interactions can be explored using
experimental data which can be collected by ATLAS and CMS
Collaborations at the LHC. These data can be analyzed through the compared
measures of some inclusive distributions and final state
correlations.

One of the aims of this paper is to carry out the  proposal for
the experimental measurements of the $Z^{0}Z^{0}$ pair correlations.

This exploration will be theoretically supported by the quantum
field theory at finite temperature $({QFT}_\beta)$ model approach [3].
It is known that the effective temperature of the vacuum or the ground state
or even the thermalized state of particles distorted by external forces is
occurring in models quantized in external fields. One of the main parameters
of the model is the temperature of the particle source under the random
source operator influence.
The main channels are the di-lepton production
$pp\rightarrow Z^{0}Z^{0}\rightarrow
2e^{-}2e^{+},~2\mu^{-}2\mu^{+},~e^{-}e^{+}\mu^{-}\mu^{+}$ in $pp$
collisions.

An efficient selection of leptons needs to be  performed
according to the following criteria. First, all leptons were
required to lie in the pseudorapidity range covered by, e.g., the CMS
muon system that is, $|\eta| \le$ 2.4. Second, the leptons were
required to be unlikely charged in pairs. Note that the
acceptances of another multipurpose detector ATLAS in the
azimuthal angle and pseudorapidity are close to the respective
parameters of CMS.

The dilepton channel is especially promising from the experimental
point of view, since it is expected that the experimental
facilities related for LHC (CMS and ATLAS detectors) will make it
possible to record muons of energy in the TeV range with a
resolution of about a few percent and an efficiency close to 100
\%. Moreover, this channel is characterized by a maximum
signal-to-background ratio in the energy region being considered.

\section{BEC in case of two particles}
\label{bec}

A pair of identical bosons with the mass $m$ produced incoherently (in ideal
nondisturbed, noninteracting cases) from an extended source will
have an enhanced probability $C_{2}(p_{1},p_{2})=
N_{12}(p_{1},p_{2})/[N_{1}(p_{1})\cdot N_{2}(p_{2})]$ to be measured
in terms of differential cross section $\sigma$, where
\begin{equation}
\label{e31}
N_{12}(p_{1},p_{2})=\frac{1}{\sigma}\frac{d^{2}\sigma}{d\Omega_{1}\,d\Omega_{2}}
\end{equation}
to be found close in 4-momentum space $\Re_{4}$ when detected
simultaneously, as compared to if they are detected separately with
\begin{equation}
\label{e32}
 N_{i}(p_{i})=\frac{1}{\sigma}\frac{d\sigma}{d\Omega_{i}}, \,\,\,
 d\Omega_{i}=\frac{d^{3}\vec p_{i}}{(2\pi)^{3}\,2E_{p_{i}}}, \,\,
 E_{p_{i}}=\sqrt {\vec p_{i}^{2}+m^{2}},\,\,\,
 i = 1, 2.
\end{equation}

On the other hand, the following relation can be used to retrieve the BEC function
$C_2(Q)$:
\begin{equation}
\label{Konstrukcia C2}
 C_2(Q) = \frac{N(Q)}{N^{ref}(Q)},
\end{equation}
where $N(Q)$ in general case refer to the numbers
for neutral gauge bosons (eg., $Z^{0} Z^{0}$) with
\begin{equation}
\label{eq_01}
 Q = \sqrt {-(p_1-p_2)_{\mu}\cdot (p_1-p_2)^{\mu}}= \sqrt{M^{2} - 4\,m^{2}}.
\end{equation}
In definitions  (\ref{Konstrukcia C2}) and  (\ref{eq_01}),  $N^{ref}$ is the number of
particle pairs without BEC and
$p_{\mu_{i}}= (\omega_{i}, \vec p_{i})$ are four-momenta of produced bosons $(i = 1,\ 2)$;
$M = \sqrt {(p_1+p_2)^{2}_{\mu}}$ is the invariant mass of the pair of
bosons.

An essential problem in extracting the correlation is the estimate of the
reference distribution $N^{ref}(Q)$ in Eq. (\ref{Konstrukcia C2}). If there are other
correlations beside the Bose-Einstein effect, the distribution $N^{ref}(Q)$ should be
replaced by a reference distribution corresponding to the two-particle distribution
in a geometry without BEC. Hence, the expression (\ref{Konstrukcia C2}) represents the ratio
between the number of $Z^{0}Z^{0}$ pairs $N(Q)$ in the real world and the reference sample
$N^{ref}(Q)$ in the imaginary world. Note that the reference sample can not be directly
observed in an experiment. Different methods are usually applied for the construction
of reference samples [1], however all of them have strong restrictions. One of the preferable methods
is to construct the reference samples directly from data. For our aim for reference sample
$N^{ref}(Q)$, it is suitable to use the pairs $ Z^{0}Z^{0}$ from different (mixed) events.

It is commonly assumed that the maximum
of two-particle BEC function $C_2(Q)$ is 2 for $\vec p_{1} =
\vec p_{2}$ if no any distortion and final state interactions are
taking into account.

There are experimental difficulties in a determination of $ Z^{0}Z^{0}$ correlations, which are
associated with acceptance limitations and limited statistics in the $ Z^{0}Z^{0}$ sample.

In general, the shape of the BEC function $C_2(Q)$ is model dependent.
The most simple form of Goldhaber-like parameterization for $C_2(Q)$
[11] has been used for data fitting:
\begin{equation}
 \label{c2_aleph}
 C_2(Q)=C_0\cdot (1+\lambda e^{-Q^2R^2})\cdot (1+\varepsilon Q) ,
\end{equation}
where $C_0$ is the normalization factor, $\lambda$ is the so-called the
chaoticity strength factor, meaning $\lambda =1$ for fully
incoherent and $\lambda =0$ for fully coherent sources; the
parameter $R$ is interpreted as a radius of the particle source,
often called as the "correlation radius", and assumed to be
spherical in this parameterization.  The
linear term in (\ref{c2_aleph}) is supposed to be account within the
long-range correlations
outside the region of BEC. Note that distribution of bosons
can be either far from isotropic, usually concentrated in some directions or
almost isotropic, and what is important that in both cases the particles
are under the
random chaotic interactions caused by other fields in the thermal
bath. In the parameterization (\ref{c2_aleph}) all of these problems
are embedded in the random chaoticity parameter $\lambda$.
To advocate the formula (\ref{c2_aleph}) it is assumed: \\
a. incoherent average over particle source where $\lambda$ serve to account for: \\
- partial coherence,\\
- long-lived resonances associated with multiple distinguishable sources,\\
- $Z^{0}Z^{0}$ purity;\\
b. spherical Gaussian density of particle emission cell (with radius $R$);\\
c. static source which means no time (energy) dependence.\\

In order to save the quantum pattern of particle production process and to avoid the static
and undistorted character of particle emitter source we also suggest to
use the $C_2(Q)$ function within $QFT_\beta$ accompanying by quantum evolution approach
in the form:
\begin{equation}
\label{c2_Kozlov}
 C_2(Q)=\xi(N)\cdot \Bigl[ 1+ \frac{1}{(1+\alpha)(1+\alpha^{\prime})}\
 \tilde\Omega(Q)+
 \frac{2{\sqrt{\alpha\alpha^{\prime}}}}
 {(1+\alpha)(1+\alpha^{\prime})}\
 \sqrt{\tilde\Omega(Q)}  \Bigr]\cdot F(Q, \Delta x),
\end{equation}
where $\xi(N)$ depends on the multiplicity $N$ as
\begin{equation}
\label{eq_02}
 \xi(N)= \frac{\langle{N (N-1)}\rangle}{\langle N\rangle^2} .
\end{equation}
The function $F(Q, \Delta x)$ that expresses the correlation magnitude as
 a function of $Q$ and two-particle relative distance $\Delta x$ is a
consequence of the Bogolyubov's principle of correlations weakening
at large distances [12]
\begin{equation}
\label{e021}
 F(Q, \Delta x) = \frac{f(Q,\Delta x)}{f(p_{1})\cdot
 f(p_{2})} = 1 + r_{f}\,Q + \ldots
\end{equation}
The function (\ref{e021}) is normalized as $F(Q, \Delta x = \infty) = 1$,
 and $r_{f}$ is the measure of correlations weakening where $r_{f} \rightarrow 0$ as $\Delta
x\rightarrow \infty $; $f(Q,\Delta x)$ is the two-particle distribution function with
$\Delta x$, while $f(p_{i})$ are one-particle probability functions with $i=1,2$.

The important parameter $\alpha$ (as well as $\alpha^{\prime})$  in
(\ref{c2_Kozlov}) summarizes our
knowledge of other than space-time characteristics of the particle
emitting source, and plays the role of a coherence parameter (see [4] for details).

The $\tilde\Omega(q)$ in (\ref{c2_Kozlov}) has the following
structure in momentum space
\begin{equation}
\label{e031}
 \tilde\Omega (Q)=\Omega (Q)\cdot\gamma (n) ,
\end{equation}
where
\begin{equation}
 \label{e0310}
 \Omega (Q) = \exp (-\Delta_{p\Re}) =
 \exp \left [-(p_1-p_2)^{\mu}\,\Re_{\mu\nu}\, (p_1-p_2)^{\nu}\right ]
\end{equation}
is the smearing smooth dimensionless generalized function,
$\Re_{\mu\nu}$ is the (nonlocal) structure tensor of the space-time
size (BEC formation domain), and it defines the spherically-like
domain of emitted (produced) bosons.

To clarify with $\gamma (n)$  in (\ref{e031}) let us emphasize that most of experiments dealing with
elementary particles at high energies are of an inclusive as one measures quantum effect
of BEC on limited samples of particles produced only. The unobserved part of the rest particle
system acts then as a kind of thermal (heat) bath influencing measured samples of data
(observables). Actually, the temperature $T$ being the most important parameter
describing the influence of such a thermal bath is occurred in this model.

The function $\gamma (n)$ reflects the quantum thermal features of BEC
pattern and is defined as
\begin{equation}
\label{e032}
 \gamma (n) = \frac{{n^2 (\bar \omega )}}{{n(\omega )\ n(\omega
 ')}} ,\ \
 n(\omega ) \equiv  n(\omega ,\beta ) =
 \frac{1}{{e^{(\omega  - \mu )\beta} - 1 }} ,\ \
 \bar\omega  = \frac{{\omega  + \omega '}}{2} ,
\end{equation}
where $n(\omega,\beta )$ is the mean value of quantum numbers for Bose-Einstein
statistics particles with the energy $\omega$ and the chemical potential $\mu$
in the thermal bath
with statistical equilibrium at the temperature $T= 1/\beta$. The
following condition $\sum_{f} n_{f}(\omega,\beta) = N$ is evident,
where the discrete index $f$ reflects the one-particle state $f$.

In terms of time-like $R_{0}$, longitudinal $R_{L}$ and transverse
$R_{T}$ components of the space-time size $R_{\mu}$ the distribution
$\Delta_{p\Re}$ looks like:
\begin{equation}
\label{e33}
 \Delta_{p\Re}\rightarrow\Delta_{pR} = (\Delta p^{0})^2 R^{2}_{0} + (\Delta p^{L})^2 R^{2}_{L} +
 (\Delta p^{T})^2 R^{2}_{T} .
\end{equation}
Seeking for simplicity one has ($R_{L}=R_{T}=R$)
\begin{equation}
\label{e34}
 \Delta_{pR} = (p^{0}_{1}-p^{0}_{2})^{2}R^{2}_{0} +
 (\vec p_{1} - \vec p_{2})^{2} \vec R^{2}
\end{equation}
for identical bosons.

 Hence, we have  introduced a new parameter $R_{\mu}$, a 4-vector, which defines the
 region of nonvanishing particle density with the space-time extension of the particle
 emission source. Expression (\ref{e0310})  must be understood in the sense that $\Omega(Q)$
 is a function that in the limit $R\rightarrow\infty$, strictly becomes a
 $\delta$-function.
For practical using with ignoring the energy-momentum dependence
of $\alpha$, and assuming that $\alpha' = \alpha$ ($\alpha$ is related with
$C_{2}(0)$ and $N$), we get the expression
with $ \Omega(Q)\simeq\exp(-Q^{2}\,R^{2})$:
\begin{equation}
C_2(Q) \simeq \xi(N) \left \{1 + \lambda_{new}(\beta)\,e^{-Q^2 R^{2}}
\left [1+\lambda_{corr}(\beta)\,e^{+Q^{2} R^{2}/2}\right ]
\right \},
\label{e58}
\end{equation}
where the new intercept function becomes as $\lambda_{new} =
\gamma(\omega,\beta)/(1+\alpha)^{2}$, and the new coherence
correction in the brackets of Eq. (\ref{e58}) carries an additional intercept function
$\lambda_{corr} = 2\,\alpha/\sqrt{\gamma(\omega,\beta)}$.
In fact, since $\alpha \neq \alpha'$
(because $\omega \neq \omega'$ and, therefore, the number
of states identified here with the number of particles $n(\omega)$ with given
energy  is also different), one can use the
general precise form  (\ref{c2_Kozlov}) for $C_2$ with details given by
Eqs. (\ref{e031}) and (\ref{e032}) and with $\alpha$ coherence function
depending on
the particle mass, the energy of final leptons produced in pairs within the decays
of $Z^{0}$'s and such characteristics of the emission process as
the temperature $T$ and
chemical potential $\mu$ occurring in the definition of $n(\omega)$ in (\ref{e032}).

Since we did not follow special assumptions on the
quantum operator level for $C_{2}(Q)$ from the initial stage, it may
correspond to a physically real and observable effect at the
LHC. This pattern may lead to a new squeezing state of
correlation region.

\section{Stochastic field and Green's function}

Let us consider the stochastic field $B_{\mu}(x)= B_{{\mu}_{\tilde{s}}}(x,\tau)$
that depends on the arbitrary random source $\tilde{s}(x)$, and the
fifth component $\tau$ means the "stochastic time". The differential
equation of an evolution of the field operator
$B_{{\mu}_{\tilde{s}}}(x,\tau)= B_{\mu}(x,\tau)$
in the system under the associated stochastic process is

\begin{eqnarray}
\label{e1}
\partial_{\tau}B_{\mu}(x,\tau)= O[B_{\mu}(x,\tau)]\, ,
\end{eqnarray}
where $O[B_{\mu}(x,\tau)]$ is the differential stochastic operator which has the form
\begin{eqnarray}
\label{e2}
 O[B_{\mu}(x,\tau)]= -\frac{1}{V}\frac{\delta J[B_{\mu}(x,\tau)]}{\delta B_{\mu}(x,\tau)} +
\tilde s_{\mu}(x,\tau)
\end{eqnarray}
with a volume $V$ being introduced by dimensional reason.
The r.h.s. of Eq. (\ref{e2}) is the so-called stochastic operator where
$J=\int d^{4} y L[B_{\mu}(y), \partial_{\nu} B_{\mu\nu}(y)]$ is the action
defined by the Lagrangian density $L$;
$\tilde s_{\mu}(x,\tau) = s_{\mu}(x,\tau) + n_{\mu} P $ carries the random
stochastic history where the memory
dissipation forces and the heat bath effects are included into
$ s_{\mu}(x) = s_{\mu}(x,\tau)$, the constant
$P$ emerges within the action of the stationary forces.  Equation (\ref{e1}) is nothing
other but the evolution equation of the Langevin type applied already
to stochastic processes on the quantum operator level in derivation of multiparticle
Bose-Einstein correlations [3].

For simplicity, we assume that $\tilde{s}_{\mu}(x)$ varies
stochastically with the Gaussian correlation function
\begin{eqnarray}
\label{e3}
\langle\tilde{s}_{\mu}(x)\tilde{s}_{\nu}(y)\rangle = const\,\delta_{\mu\nu}
\exp (-z^{2}/l^{2}_{ch})\, ,
\end{eqnarray}
where $z_{\nu}=(x-y)_{\nu}$, and $const$ is the strength of the noise
described by the distribution function $\exp (-z^{2}/l^{2}_{ch})$ with
$l_{ch}$ being the noise characteristic scale.
Both $const$ and $\l_{ch}$ define the influence of the (Gaussian) noise
on, e.g., correlations between particles that "feel" an action of
an environment. Actually, Eq. (\ref{e1}) can be transferred to the standard field
equation of motion (in Euclidean space)
\begin{eqnarray}
\label{e4}
\frac{1}{V}\frac{\delta J[B_{\mu}(x)]}{\delta B_{\mu}(x)} = \tilde s_{\mu}(x)
\end{eqnarray}
with the source $\tilde s_{\mu}(x)$ if both $B_{\mu}$ and $\tilde s_{\mu}$ do not
depend on "stochastic time" $\tau$.
In classical theory, the random process given by
$\tilde {s}_{\mu}(x)$ is nothing other but the white (Gaussian) noise.

In this paper, we  focus on the role of  particle masses and energies,
effects of coherence and distortion, and the heat bath influences which are rather
important to describe the correlations between particles. To solve this problem,
especially to derive the memory term in evolution equation  one can use the
general properties of  ${QFT}_{\beta}$. The model is defined by the following generating
functional in four-dimensional space-time
\begin{eqnarray}
\label{e5}
Z = \int D B_{\mu}\exp \left [ -i\int d^{4} x L(B_{\mu},B_{\mu\nu})\right ]\, ,
\end{eqnarray}
where
\begin{eqnarray}
\label{e6}
L = -\frac{1}{4} B_{\mu\nu}B^{\mu\nu} + \frac{1}{2}(m^{2} + U)B_{\mu}B^{\mu}
\end{eqnarray}
with $B_{\mu\nu} = \partial_{\mu}B_{\nu} - \partial_{\nu}B_{\mu}$.

The direct calculations using the solution of Eq. (\ref{e1}) with the Lagrangian density
(\ref{e6}) leads to the propagator of the field $B_{\mu}(x,\tau)$ distorted by
$\tilde s_{\mu}(x,\tau)$. The transverse part of $B_{\mu}(x,\tau)$ will give the
correct expression for the Euclidean vector field propagator at $\tau\rightarrow\infty$.

We are working with fields that correspond to a thermal
field $B_{\mu}(x)$ with the standard definition of the Fourier transformed propagator
$F[\tilde G_{\mu\nu}(p)]$
\begin{equation}
F[\tilde G_{\mu\nu}(p)]= G_{\mu\nu}(x-y) =
Tr\left\{T[B_{\mu}(x)B_{\nu}(y)]\rho_{\beta}\right\}
\label{e7}
\end{equation}
with $\rho_{\beta}= e^{-\beta H}/Tr e^{-\beta H}$ being the density matrix of a local
system in equilibrium at temperature $T$ under the Hamiltonian $H$
\begin{equation}
H = \int \frac{d^{3}\vec p}{(2\pi)^{3} 2p^{0}}\,p^{0}
\sum_{\lambda =1}^{3} b^{{\lambda}^{+}}(p)b^{\lambda}(p)
\label{e8}
\end{equation}
with the operators of annihilation $b^{\lambda}(p)$ and creation $b^{{\lambda}^{+}}(p)$
to be defined later.

The interaction of $B_{\mu}(x)$ with the external field is given by the potential $U$.
The equation of motion is
\begin{equation}
(\nabla^2 + m^2)B_{\mu}(x) = -J_{\mu}(x),
\label{e9}
\end{equation}
where $J_{\mu}(x) = U B_{\mu}(x)$ is the source density operator.
A simple model like this allows one
to investigate the origin of the
unstable state
of the thermalized equilibrium in a nonhomogeneous external field under the influence
of  source density operator $J_{\mu}(x) = U B_{\mu}(x)$. For example, the source
can be considered
as $\delta$-like generalized function, $J_{\mu}(x)=\tilde\mu\,\rho(x,\epsilon)B_{\mu}(x)$
in which
$\rho(x,\epsilon)$ is a $\delta$-like succession giving the $\delta$-function as
$\epsilon\rightarrow 0$ (where $\tilde\mu$ is some massive parameter). This model is useful
because the $\delta$-like potential $U(x)$ provides the conditions
for  restricting the particle emission domain (or the deconfinement region). We suggest the
following form:
\begin{equation}
 J_{\mu}(x) = - J_{sys}(x)\,B_{\mu}(x) + J_{R_{\mu}}(x),
\label{e10}
\end{equation}
where the source $J_{\mu}(x)$ is a sum of a regular systematic motion part $J_{sys}(x)$
and the random source $J_{R_{\mu}}(x)$. The
equation of motion (\ref{e9}) becomes
\begin{equation}
 [\nabla^2 + m^2 - J_{sys}]B_{\mu}(x) = -J_{R_{\mu}}(x),
\label{e11}
\end{equation}
and the propagator satisfies the following equation (in the Fourier transformed
form labeled by tilde):
\begin{equation}
[p^{2}_{\mu} - m^2 +\tilde J_{sys}]\tilde G_{\mu\nu}(p_{\mu}) = \tilde d_{\mu\nu} (p),
\label{e12}
\end{equation}
where
\begin{equation}
d_{\mu\nu}(x) = \left (g_{\mu\nu} + \frac{1}{m^{2}}\frac{\partial^{2}}
{\partial x_{\mu}\partial x_{\nu}}\right )\delta (x).
\label{e13}
\end{equation}
As the standard point, the Green's function of the vector field can be obtained from the one
of the scalar field
acting by the relevant operator
$g_{\mu\nu} + m^{-2}\partial^{2} / (\partial x_{\mu}\partial x_{\nu})$.

The solution of Eq. (\ref{e9}) is
\begin{equation}
B_{\mu}(x) = -\int dy\, G_{\mu\nu} (x,y)\,J_{R_{\nu}}(y),
\label{e14}
\end{equation}
where the Green's function obeys the Eq. (\ref{e12}).
\section{ Green's function and kernel operator}

Let us go to the thermal field operator $B_{\mu}(x)$ by means of the linear combination
of the frequency parts $B_{\mu}^{1}(x)$ and $B_{\mu}^{{2}^{+}}(x)$
\begin{equation}
B_{\mu}(x) = B_{\mu}^{1}(x) + B_{\mu}^{{2}^{+}}(x)
\label{e24}
\end{equation}
with $[B_{\mu}(x), B_{\nu}(y)] = i D_{\mu\nu}(x-y)$ and

$$ B_{\mu}^{1} (x) = \int \frac{d^{3}\vec{p}}{(2\pi)^{3} 2 (\vec p^{2} +m^{2})^{1/2}}
\sum_{\lambda =1}^{3} \epsilon_{\mu}^{(\lambda)}(p)\tilde b^{(\lambda)}(p)\,e^{-ipx}, $$
$$B_{\mu}^{{2}^{+}}(x) = \int \frac{d^{3}\vec{p}}{(2\pi)^{3} 2 (\vec p^{2} +m^{2})^{1/2}}
\sum_{\lambda =1}^{3} \epsilon_{\mu}^{(\lambda)}(p)\tilde b^{{(\lambda)}^{+}}(p)\,e^{ipx}. $$
The following properties of polarization vectors are the standard ones:
$$\epsilon_{\mu}^{(\lambda)}(p)\epsilon_{\mu}^{(\lambda^{\prime})}(p) = g_{\lambda\lambda^{\prime}},$$
$$\sum_{\lambda =1}^{3} \epsilon_{\mu}^{(\lambda)}(p)\epsilon_{\nu}^{(\lambda)}(p) =
- g_{\mu\nu} + \frac{p_{\mu}p_{\nu}}{m^{2}}.$$
We assume that the deviation from the asymptotic free state given by the operator $a(\vec p,t)$
is provided by the random operator $r(\vec p,t):a(\vec p,t)\rightarrow
b(\vec p,t) = a(\vec p,t) + r(\vec p,t)$.
The operators $\tilde b^{(\lambda)}(p)$ and $\tilde b^{{(\lambda)}^{+}}(p)$ obey the following
equations in $\Re_{4}$ (see details in [3]):
\begin{equation}
[\omega - \tilde K(p)]\tilde b^{(\lambda)}(p) = \tilde F(p) + \rho(\omega_{P},\epsilon),
\label{e27}
\end{equation}
\begin{equation}
[\omega - \tilde K^{+}(p)] \tilde b^{{(\lambda)}^{+}}(p) =
\tilde F^{+}(p) + \rho^{*}(\omega_{P},\epsilon),
\label{e28}
\end{equation}
where $p_{\mu} = (\omega = p^{0}, \vec p)$. Both equations (\ref{e27}) and (\ref{e28}) can be
transformed into new equations for the frequency parts
$ B_{\mu}^{1}(x)$ and $B_{\mu}^{{2}^{+}} (x)$
\begin{equation}
i\partial_{0} B_{\mu}^{1} (x) + \int_{\Re_{4}} K(x-y)\,B_{\mu}^{1}(y)dy = f_{\mu}(x),
\label{e29}
\end{equation}
\begin{equation}
- i\partial_{0} B_{\mu}^{{2}^{+}} (x) + \int_{\Re_{4}} K^{+}(x-y)\,B_{\mu}^{{2}^{+}}(y)dy =
f_{\mu}^{+}(x),
\label{e30}
\end{equation}
where
\begin{equation}
f_{\mu}(x) = \int \frac{d^{3}\vec{p}}{(2\pi)^{3} 2 (\vec p^{2} +m^{2})^{1/2}}
\sum_{\lambda =1}^{3} \epsilon_{\mu}^{(\lambda)}(p)
[\tilde F(p) + \rho(\omega_{P},\epsilon)]e^{-ipx},
\label{e31}
\end{equation}
\begin{equation}
f_{\mu}^{+}(x) = \int \frac{d^{3}\vec{p}}{(2\pi)^{3} 2 (\vec p^{2} +m^{2})^{1/2}}
\sum_{\lambda =1}^{3} \epsilon_{\mu}^{(\lambda)}(p)
[\tilde F^{+}(p) + \rho^{*}(\omega_{P},\epsilon)]e^{ipx}.
\label{e32}
\end{equation}
The equations for field components $B_{\mu}^{1} (x)$ and  $B_{\mu}^{{2}^{+}}(x)$
(\ref{e29}) and (\ref{e30}), respectively, are nonlocal
within the presence of the  formfactors $K(x-y)$ and $K^{+}(x-y)$, respectively.
In principle, these formfactors can
admit the description of locality for nonlocal interactions.
At this stage, it must be stressed that we have new generalized evolution
equations (\ref{e29}) and (\ref{e30}), which retain  the general
features of the propagating and
interacting of the quantum vector fields with mass $m$ that are in the heat bath
(thermal reservoir)
and are chaotically distorted by  other fields. For  further analysis,
let us rewrite the system of Eqs. (\ref{e29}) and (\ref{e30}) in the following form:
\begin{equation}
i\partial_{0} B_{\mu}^{1} (x) +  K(x)\star B_{\mu}^{1} (x) = f_{\mu}(x),
\label{e34}
\end{equation}
\begin{equation}
- i\partial_{0} B_{\mu}^{{2}^{+}} (x) +  K^{+}(x)\star B_{\mu}^{{2}^{+}} (x) = f_{\mu}^{+}(x),
\label{e35}
\end{equation}
where $A(x)\star B(x)$ is the convoluted function of the generalized functions
$A(x)$ and $B(x)$.
Applying the direct Fourier transformation to both sides of Eqs.
(\ref{e34}) and (\ref{e35}) with the following properties of the
Fourier transformation
$$ F[K(x)\star B_{\mu}^{i} (x) ] = F[K(x)]\,F[B_{\mu}^{i} (x)] \,\,\, (i = 1, 2^{+}),$$
we  get two equations
\begin{equation}
[p^{0} + \tilde K(p)]\tilde B_{\mu}^{1}(p) = F[f_{\mu}(x)],
\label{e38}
\end{equation}
\begin{equation}
[- p^{0} - \tilde K^{+}(p)]\tilde B_{\mu}^{{2}^{+}}(p) = F[f_{\mu}^{+}(x)].
\label{e39}
\end{equation}
Finally, we have got the following equation for $\tilde B_{\mu}(p)$ field:

\begin{equation}
[- p^{0} +  \tilde K^{+}(p)][p^{0} + \tilde K(p)]\tilde B_{\mu}(p) = \tilde T_{\mu}(p),
\label{e040}
\end{equation}
where
$$ \tilde T_{\mu}(p) = [- p^{0} + \tilde K^{+}(p)]F[f_{\mu}(x)]+
 [p^{0} + \tilde K(p)]F[f_{\mu}^{+}(x)]. $$
We are now  at the stage of the main strategy:  one has to identify the field
$B_{\mu} (x)$ and the random source operator $J_{{R_{\mu}}}(x)$,
introduced in Eq. (\ref{e11}, with the Fourier transformed field
$\tilde B_{\mu}(p)$  and $\tilde T(p)$ in (\ref{e040}), respectively.

The next step is our requirement that Green's function
$\tilde G_{\mu\nu}(p)$ in Eq. (\ref{e12}) and  the function
$ \Gamma_{\mu\nu}(p)$, satisfying the equation
\begin{equation}
[- p^{0} +  \tilde K^{+}(p)][p^{0} + \tilde K(p)]\tilde \Gamma_{\mu\nu}(p) = g_{\mu\nu}
\label{e42}
\end{equation}
must be equal to each other, i.e.
$$F[\tilde G_{\mu\nu}(p) - \tilde \Gamma_{\mu\nu}(p)] = 0.$$

The kernel operator $\tilde K(p)$ is
\begin{equation}
\tilde K(p) \simeq \epsilon \sqrt {1 + \frac{m^{2}}{\epsilon^{2}}},
\label{e43}
\end{equation}
where $\epsilon = 2 \sqrt {\vec k_{l}^{2} + m_{l}^{2}}$ is the total energy
of the final lepton-antilepton pair (with momentum $\vec k_{l}$ and the mass
$m_{l}$ for the lepton)
produced within the decay of $Z^{0}$ boson being in the rest frame.
To get $\tilde K (p)$ in the form( \ref{e43}) we used the fact that the full
Green's function $\tilde G_{\mu\nu}(p)$ is given by the corresponding full
Green's function of the scalar field [13] under the action by the differential
operator $(g_{\mu\nu} - m^{-2}\, p_{\mu}\, p_{\nu})$.

\section {Source size}
It has been emphasized [4] that
there are two different scale parameters in the model considered
here. One of them is the so-called "correlation radius" $R$ introduced
in (\ref{c2_aleph}) and (\ref{c2_Kozlov}) with (\ref{e33}). In fact,
this $R$-parameter gives the pure size of the particle emission
source without the external distortion and interaction coming from
other fields. The other (scale) parameter is the stochastic
scale $L_{st}$ which carries  the dependence of the particle mass,
the $\alpha$-coherence degree and what is very important --- the
temperature $T$-dependence:
\begin{equation}
\label{e36}
 L_{st}\simeq {\left[\frac{1}{\alpha(N)\, {\vert p^{0}-\tilde K(p)\vert}^{2}\,
 n(m,\beta)}\right ]}^{\frac{1}{2}}\rightarrow
 {\left[\frac{1}{\alpha(N)\, 4\vec k^{2}_{l}\,
 {\vert 1-\delta_{k}\vert}^{2}\,\bar n(m,\beta)}\right ]}^{\frac{1}{5}},
\end{equation}
where
$$\delta_{k} = \sqrt {1 + \frac{m^{2}}{4\vec k^{2}_{l}}}$$
and the lepton mass $m_{l}$ is neglected.

It turns out that the scale $L_{st}$ defines the range of stochastic forces.
 This effect is given by $\alpha
(N)$-coherence degree which can be estimated from the experiment within
the two-particle BEC function $C_{2}(Q)$ when $Q$ close to zero, $C_{2}(0)$,
at fixed value of mean multiplicity $\langle N\rangle$:
\begin{equation}
 \label{e37}
\alpha (N)\simeq \frac{2-\bar C_{2}(0) + \sqrt {2-\bar
C_{2}(0)}}{\bar C_{2}(0)-1},\,\,\, \bar C_{2}(0)= C_{2}(0)/\xi(N) .
\end{equation}
In formula (\ref{e36}),  $\bar n(m,\beta)$ is the thermal
relativistic particle number density
\begin{equation}
 \label{e38}
\bar n(m,\beta) = 3\int\frac{d^{3}\vec
p}{(2\,\pi)^{3}}\,n(\omega,\beta)=3\frac{\mu^2 +m^2}
{2\,\pi^2}\,T\,\sum_{l=1}^{\infty}\frac{1}{l}K_{2}\left
(\frac{l}{T}\sqrt{\mu^2 + m^2}\right) ,
\end{equation}
where $K_{2}(...)$ is the modified Bessel function.

The coherence function $\alpha$ is another very important one that
summarizes our knowledge of other than space-time characteristics
of the particle emission source, and the prediction of $\alpha$ from an
experiment is very instructive aim itself. For $\alpha =0$, one actually finds
$$ 1 < C_{2}(Q) < \xi (N) (1 + \gamma e^{-Q^{2}R^{2}})$$
which is nothing other but the Goldhaber parameterization [11] with
$0< \gamma < 1$ being a free parameter adjusting the observed value of $C_{2}(Q=0)$.

Within our aim to
explore the correlation between $Z^{0}Z^{0}$ the scale $L_{st}$ has the form
\begin{equation}
 \label{e39}
L_{st}\simeq {\left [\frac{e^{\sqrt{m^{2}+\mu^{2}}/T}}{12\, \alpha(N)\,\vec k^{2}_{l}\,
{(m^{2}+\mu^{2})}^{3/4}{\left
(\frac{T}{2\,\pi}\right)}^{3/2}\, \left
(1+\frac{15}{8}\frac{T}{\sqrt{m^{2}+\mu^{2}}}\right )
{\vert 1-\delta_{k}\vert}^{2}}\right ]}^{\frac{1}{5}} ,
\end{equation}
where the condition $l\,\beta\,\sqrt{m^{2}+\mu^{2}} >1$  for any integer $l$
in (\ref{e38}) was taken into account. The only lower temperatures will drive $L_{st}$
within formula (\ref{e39}) even if $\mu = 0$ and $l=1$ with the condition $T < m$.

Note that the
condition $\mu < m$ is a general restriction in the relativistic
"Bose-like gas", and $\mu = m$ corresponds to the Bose-Einstein
condensation.

For high enough $T$ no $\mu$ - dependence is found for $L_{st}$:
\begin{equation}
 \label{e40}
L_{st}\simeq {\left [\frac{\pi^{2}}{12\,\zeta (3)\, \alpha(N)\,\vec k^{2}_{l}\,
T^{3} {\vert 1-\delta_{k}\vert}^{2}}\right ]}^{\frac{1}{5}} ,
\end{equation}
where the condition $T > l \sqrt{m^{2}+\mu^{2}}, l=1,2, ...$ is taken into account.
The origin of formula (\ref{e40}) comes from
\begin{equation}
 \label{e41}
\bar n(m,\beta)\rightarrow \bar n(\beta)\simeq \frac{3\,T^{3}}{\pi^{2}}\,\zeta (3)
\end{equation}
where neither a $Z^{0}$ boson mass nor the $\mu$ - dependence occurred;
$\zeta (3) = \sum^{\infty}_{l=1}l^{-3} = 1.202$ is the zeta-function with the
argument $3$.

To be close to the experiment  there is necessary to include transverse momenta,
where the $Z^{0}$ boson mass $m$, in
Eqs. (\ref{e38}), (\ref{e39}), (\ref{e40}) is replaced by the transverse mass
$m_T = \sqrt{m^{2} + p^{2}_{T}}$.

Actually, the increasing of $T$ leads to squeezing of the domain of
stochastic force influence, and $L_{st}(T=T_{0})= R$ at some effective temperature $T_{0}$.
The higher temperatures, $T > T_{0}$,
satisfy to more squeezing effect and at the critical temperature
$T_{c}$ the scale $L_{st}(T=T_{c})$ takes its minimal value.
Obviously $T_{c}\sim O(200~GeV)$ defines the phase transition where
the chiral symmetry restoration will occur.
Since in this phase all
the masses tend to zero  and $\alpha\rightarrow 0$ at $T>T_{c}$ one
should expect the sharp expansion of the region with
$L_{st}(T>T_{c})\rightarrow \infty$.

The qualitative relation between $R$ and $L_{st}$ above mentioned is the only one we
can emphasize in order to explain the mass dependence of the source size.

\section{Conclusions}

To summarize: the theoretical proposal for two-particle Bose-Einstein correlation
function in case of $Z^{0}Z^{0}$ pairs in $pp$ collisions is carried out for the
first time.

The correlations of two bosons in 4-momentum space presented in this paper offer useful
and instructive complimentary viewpoints to theoretical and experimental works in
multiparticle femtoscopy and interferometry measurements at hadron colliders.

We find the time dependence of correlation function calculated in time-dependent external
field provided by the operator $r(\vec p, t)$ and the chaotic coherence function
$\alpha(m,\beta)$. The result can be compared with the static correlation functions (see,
e.g., [14] and the references therein mainly devoted to heavy-ion collisions) and
also can be used for experimental data fitting.

The stochastic scale $L_{st}$ decreases with increasing temperatures slowly at low
temperatures, and it decreases rather abruptly when the critical temperature is approached.

Our results first predicted for correlation radius $R$ are both$ Z^{0}$
boson mass and lepton energy dependent
\begin{equation}
 \label{e411}
R \sim \frac{e^{m/{5T_{0}}}}{\alpha ^{1/5}\,{\vert\vec k_{l}\vert}^{2/5}
\,m^{3/10}\, T_{0}^{3/10}}
 \end{equation}
for low values of $T_{0} < m$, while for higher temperatures,
$T_{0} > \sqrt {m^{2} + \mu^{2}}$, one has
\begin{equation}
 \label{e412}
R \sim \frac{1}{\alpha ^{1/5}\, {\vert\vec k_{l}\vert}^{2/5}\, T_{0}^{3/5}}.
 \end{equation}
The theoretical correlation radius $R$ at temperature $T_{0}$ decreases as
$Z^{0}$-boson momentum increases.
Both estimations (\ref{e411}) and (\ref{e412}) serve as the first approximation
to explain the experimental data at different $\sqrt s$ and hence at $T$.
We claim that the experimental measuring of $R$ (in $fm$) can provide the precise estimation
of the effective temperature $T_{0}$ which is the main thermal character
in the $Z^{0}Z^{0}$ pair emitter source (given by the effective dimension $R$)
in the proper leptonic decaying channel $Z^{0}Z^{0}\rightarrow l\bar ll\bar l$
with the final lepton energy $\sqrt{\vec k^{2}_{l} + m^{2}_{l}}$ at
given $\alpha$ fixed by $C_{2}(Q=0)$ and $\langle N\rangle$. Actually, $T_{0}$ is
the true temperature in the region of multiparticle production with dimension
$R = L_{st}$, because at this temperature it is exactly the creation of two
particles ($Z^{0}Z^{0}$) occurred, and these particles obey the criterion of BEC.


\end{document}